\documentclass{PoS}

\usepackage{overpic}

\title{A Search for Dark Matter from Dwarf Galaxies using VERITAS}

\ShortTitle{Search for Dark Matter from Dwarf Galaxies using VERITAS}

\author{\speaker{Benjamin Zitzer} for the VERITAS Collaboration\thanks{veritas@veritas.sao.arizona.edu}\\
        McGill University\\
        E-mail: \email{bzitzer@mcgill.physics.ca}}

\abstract{In the cosmological paradigm, cold dark matter (DM) dominates the mass content of the Universe and is present at every scale. Candidates for DM include many extensions of the standard model, such as weakly interacting massive particles (WIMPs) in the mass range from $\sim$10 GeV to greater than 10 TeV. The self-annihilation or decay of WIMPs in astrophysical regions of high DM density can produce secondary particles including very high energy (VHE) gamma rays with energy up to the DM particle mass. VERITAS, an array of atmospheric Cherenkov telescopes, sensitive to VHE gamma rays in the 85 GeV-30 TeV energy range, has been utilized for the search for this DM signature. The astrophysical objects considered to be candidates for indirect DM detection by VERITAS are dwarf spheroidal galaxies (dSphs) of the Local Group and the Galactic Center, among others. This presentation reports on the observations of five dSphs, and the results from a joint DM search from these objects.}

\FullConference{The 34th International Cosmic Ray Conference,\\
		30 July- 6 August, 2015\\
		The Hague, The Netherlands}

\begin{document}

\section{Introduction}
The search for the existence of the particle candidates comprising dark matter (DM) is still an open question. Many theories for DM involve extensions of the standard model (SM) such as supersymmetry (SUSY) \cite{1996PhR...267..195J} or theories with extra dimensions \cite{2003NuPhB.650..391S} with weakly interacting massive particles (WIMPs) that would annihilate or decay into standard model particles, most producing a continuum of $\gamma$ rays with energies up to the DM particle mass \cite{bertone2009} or mono-energetic $\gamma$-ray lines \cite{bringmann2008}. The indirect search for SM particles from astrophysical objects with a large inferred DM density is a important complement to direct searches for DM interactions and accelerator production experiments. 

Attractive objects for indirect DM searches are nearby massive objects with high inferred DM density which are not known or expected to be $\gamma$-ray emitters. The Galactic Center is expected to be the brightest source of $\gamma$ rays resulting from DM annihilation; however VHE emission exists coincident with the position of SgrA* \cite{HESS_SgrA} and VHE emission along the Galactic Center ridge \cite{HESS_SgrA_Diffuse}\cite{ASmithICRC2015}, makes DM searches in the Galactic center a complicated, but not impossible, prospect. Other targets for indirect DM searches are \textit{Fermi}-LAT targets without counterparts at other wavelengths, which are potentially DM Galactic sub-halos \cite{2010PhRvD..82f3501B}\cite{NeitoICRC2015} and galaxy clusters \cite{2012ApJ...757..123A}. Finally, dwarf spheroidal galaxies (dSphs) are relatively close ($\sim$50 kpc), and have a low rate of active star formation, suggesting a low background from conventional astrophysical VHE processes \cite{2000PhRvD..61b3514B}. This work will focus on the results of observations of dwarf galaxies with the Very Energetic Radiation Imaging Telescope Array System (VERITAS).  

\section{VERITAS Data and Analysis}

VERITAS has devoted a significant portion ($\sim$15$\%$ in recent years of the total dark time observing budget) to the observation of indirect DM targets, with most of that time on dSphs. This time is divided between the `classical' dSphs, such as Draco and Ursa Minor (UMi), which have large stellar populations and therefore a well-constrained DM density profile, and the `ultra-faint' dSphs, mostly Segue 1, which has a large inferred DM luminosity because of its close distance, but a lower stellar population and therefore a larger uncertainty in the DM luminosity. The inferred DM luminosity is referred to as the `J factor' or `astrophysical factor'  and is defined as for annihilation:  

\begin{equation}
J_{ANN} = \int\int\rho(l,\Omega)^{2}dld\Omega
\end{equation}
where $\rho(l,\Omega)$ is the DM density profile, $\Omega$ is the solid angle around the dSph and $l$ is the distance along the line of sight. VERITAS divides its time between the classical and ultra-faint dSphs to give a larger coverage in RA bands and as a safeguard in case any one particular dSph falls out of favor of being a good DM target. As an example, the ultra-faint dSph Segue 2 was speculated to have a large J factor because of its relatively close-by location ($\sim$35 kpc) \cite{2009MNRAS.397.1748B} indicating a good DM target, then additional studies led to a re-evaluation and a much lower J factor. \cite{AlexJProfile}. 

During the period from the beginning of four-telescope operations in 2007 to summer 2013, $\sim$230 hours of quality data of dSphs have been taken. Quality data for this work requires nominal hardware operations of all four telescopes, clear weather, and low night-sky background (NSB) in order to get the lowest possible energy threshold. The data span three VERITAS epochs: before and after the T1 relocation \cite{2009arXiv0912.3841P} and after the upgrade of the array with FPGA-based topological triggers \cite{2013arXiv1307.8360Z} and camera pixels with higher quantum-efficiency photomultiplier tubes (PMTs) \cite{2013arXiv1308.4849D}. The five dSphs used in this study are Segue 1, Ursa Minor (UMi), Draco, Bo{\"o}tes 1, and Willman 1. The observations and results of the data analysis of these dSphs are summarized in Table 1.

Data analysis mostly uses the standard VERITAS techniques \cite{2008ICRC....3.1385C}, with a few notable exceptions. In order to construct a viable background spectrum for each observation run (which is required by the event weighting method described later in the text) the cosmic-ray background was subtracted using an annulus centered around the telescope array tracking position as opposed to the target location  \cite{2013arXiv1307.8367Z}. Also, in this work, looser cuts optimized for soft spectral sources were used, in order to obtain the lowest possible energy threshold. The combination of deep exposures and the lower energy threshold revealed background systematics in the cameras, namely a gradient along the direction of zenith angle of observation and deficit regions in the camera due to optically bright stars. The 3.8 magnitude star $\eta$-Leonis located 0.68$^{\circ}$ away from Segue 1 was of particular concern. The zenith gradient was corrected for by using a zenith-dependent acceptance function \cite{2015arXiv150300743B}\cite{2003AA...410..389R}. The bright stars were corrected for using a 2D elliptical fit to each shower image, as opposed to a moment analysis \cite{2012AIPC.1505..709C}. This has the added benefit of interpolating over camera pixels that are shut off due to high currents or have raised cleaning thresholds due to higher NSB levels \cite{2015arXiv150300743B}, and improving the PSF, particularly at higher energies.   

\begin{table}[t]
	\begin{center}
	\begin{tabular}{ l | c | c | c | c }
	Dwarf & Live time  & $\log_{10}J$ &Significance &  $F_{-12}^{95\%}$   \\
	 & [hrs] & [GeV$^{2}$cm$^{-5}$] &[$\sigma$] & [$10^{-12} {\mathrm{cm}}^{-2} {\mathrm{s}}^{-1}$] \\
	\hline
	Segue 1   	 &   92.0  & 19.4$^{+0.3}_{-0.4}$	&    0.7	   	& 0.34\\
	Ursa Minor  &  60.4  & 18.9$^{+0.3}_{-0.2}$	&   -0.1	   	& 0.37\\
	Draco       	 &  49.8   & 18.8$\pm0.1$ 			&   -1.0  		& 0.15 \\
	Bo{\"o}tes  &  14.0   & 18.2$\pm0.4$ 			&   -1.0 		& 0.40\\
	Willman 1   &  13.7   &   N/A				&   -0.6    		& 0.39\\
	\end{tabular}
	\caption{Summary of dSph data used in this work. $J$ factors are from \cite{AlexJProfile} with an integration radius of 0.5$^{\circ}$ around each Dwarf. Dwarf galaxy detection significance is calculated from Li \& Ma equation 17 \cite{1983ApJ...272..317L} and the integral flux upper limits at 95\% confidence flux is in units of $10^{-12} {\mathrm{cm}}^{-2} {\mathrm{s}}^{-1}$ above 300 GeV, assuming a spectral index of -2.4, using the Rolke method with bounded intervals \cite{2005NIMPA.551..493R}. }
	\label{tab:dwarfsresult}
	\end{center}
\end{table} 
\section{Dark Matter Search}

To search for dark matter in all five dSphs simultaneously, an event weighting method is employed. The full details of the method are in \cite{2015PhRvD..91h3535G}; it is only summarized here. It improves on standard IACT techniques for DM searches by utilizing the individual energies and spatial properties of the individual events, as well as the instrumental and astrophysical background properties. Each event in the region-of-interest (defined as the ON region in most IACT analysis) is assigned a weight, $w_{i}$, based on the dwarf field it originates from, its reconstructed energy $E$, its reconstructed angular distance away from the dwarf, $\theta$ and dwarf FOV it originated from, $\nu$. The test statistic for the hypothesis of $\gamma$ rays being produced by dark matter is the sum of the weights from all events: $T = \sum w_{i}$. The index $i$ is summed over all ON events for all dSphs. The most optimal form of the weight function is
\begin{equation}
w(\nu,E,\theta) = \log{\left[1 + \frac{s(\nu,E,\theta)}{b(\nu,E,\theta)}\right]}
\end{equation}
where $s(\nu,E,\theta)$ and $b(\nu,E,\theta)$ are the total number of expected signal and background events with those properties, respectively. The number of expected background events is extracted by a spectrum generated by the OFF data for each data run using the Crescent method described in the previous section. The background is assumed to be isotropic within this region, so $b(\nu,E,\theta)$ is proportional to $\sin(\theta)d\theta$.   

The expected signal $s(\nu,E,\theta)$ is determined by a convolution of the dark matter annihilation flux with the VERITAS instrument response
\begin{equation}
s(\nu,E,\theta) = \frac{dN(\nu,E,\theta)}{dEd\Omega}dE2\pi\sin(\theta)d\theta
\end{equation}
where the number of events reconstructed with energy $E$ and angular separation $\theta$ is given by the convolution
\begin{equation}
\frac{dN(\nu,E,\theta)}{dEd\Omega} = \int_{E_t}\int_{\Omega_t}dE_{t}d\Omega_{t}\frac{F(E_{t},\theta_{t})}{dE_{t}d\Omega_{t}}R(E,\theta | E_{t},\theta_{t})
\end{equation}
where the subscript $t$ denotes true energies and directions and the function $R$ is the VERITAS response which is a function of the effective area, live time, instrument PSF, and energy dispersion (the probability $E$ given $E_{t}$). Effective area, PSF and energy dispersion are generated from $\gamma$-ray simulations. The function $F$ is the $\gamma$-ray flux from dark matter annihilation:
\begin{equation}
 \frac{F(E_{t},\theta_{t})}{dE_{t}d\Omega_{t}} = \frac{\langle\sigma\nu\rangle}{8\pi M^{2}}\frac{dN_{\gamma}(E)}{dE}\frac{dJ(\theta)}{d\Omega}
\end{equation}
where $dN_{\gamma}(E)/dE$ is the spectrum of gamma-rays produced per annihilation, $dJ(\theta)/d\Omega$ is the DM density profile, $M$ is the dark matter particle mass and $\langle\sigma\nu\rangle$ is the velocity-weighted annihilation cross section. The single annihilation spectrum used in this work is from PPP4 \cite{PPP4} and the DM density profile is from \cite{AlexJProfile}. An example of the form of $w_{i}$ as a function of $E$ and $\theta$ for a dark matter particle of mass 10 TeV is shown in Figure 1.  

\begin{figure}
	\centering
	\includegraphics[scale=0.4]{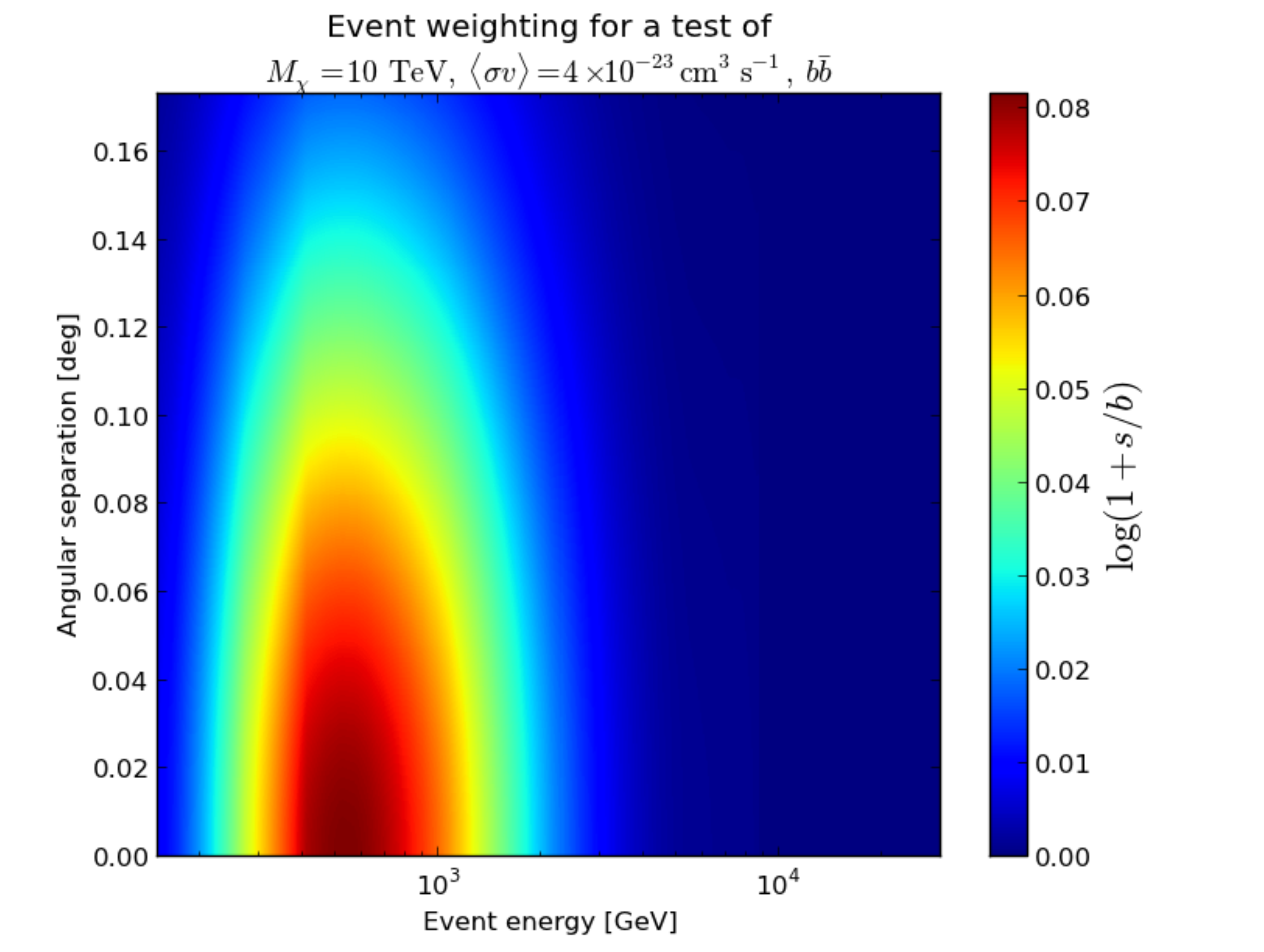}
	\caption{Example of the form of $w_{i}$ for a dark matter particle of mass of 10 TeV as a function of reconstructed energy $E$ and reconstructed angular distance from the dwarf $\theta$.}
\end{figure}

In order to search for annihilation or set limits on the cross section, we need to compute a probability distribution for $T$, with the null hypothesis being that the observed test statistic, $T_{obs}$, is due to only background processes. The detection significance is defined as the probability that $T$ is less than $T_{obs}$ under the background-only hypothesis (equivalent to  $\langle\sigma\nu\rangle$=0). The description of computing the probability distributions is detailed in \cite{2015PhRvD..91h3535G}. This distribution is computed for specified values of $M$ and choices of DM annihilation channels. 
 
The search for annihilation is performed by measuring $T_{obs}$ and comparing this with the probability distribution for $T$ due to background. A search for a individual dwarf galaxy is performed by setting weights from all other dSphs to 0. The search for dark matter is performed as a function of mass and with the consideration of a 100\% branching ratio to heavy quarks ($b\bar{b}$), leptons ($\tau^{+}\tau^{-}$) and a two-photon final state. Figure 2 shows the combined search for four of the five dSphs in this study, using the data summarized in Table 1. Willman 1 is not used because the DM profile could not be reliably determined \cite{Willman2011}.  Figure 2 shows a non-detection of DM in the mass range from 100 GeV to 100 TeV at a significance level not exceeding the 2$\sigma$ level above the background. 

\begin{figure}
	\centering
	\begin{overpic}[scale=0.55]{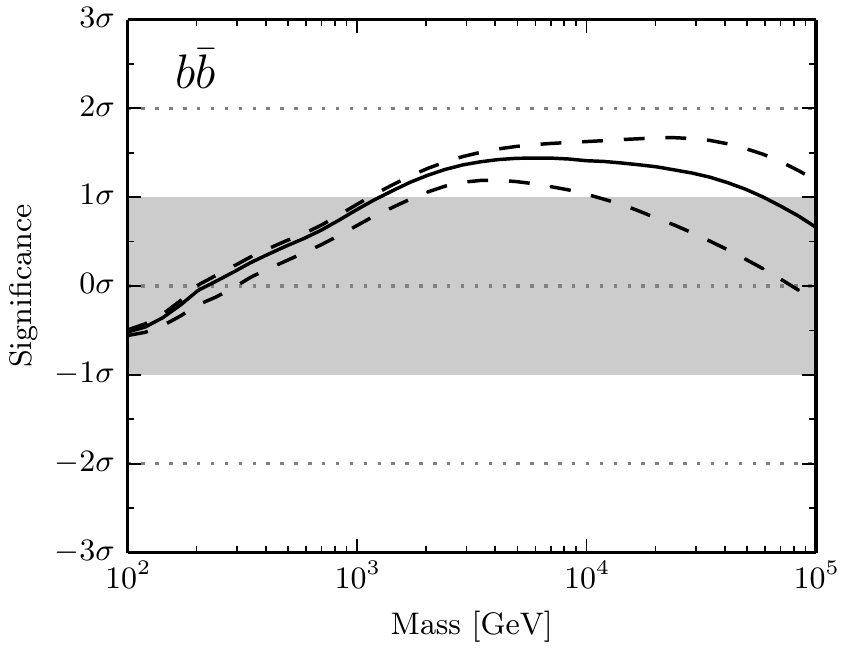}
	\put(20,15){\small\color{red} VERITAS - ICRC 2015}
	\end{overpic}
	\begin{overpic}[scale=0.55]{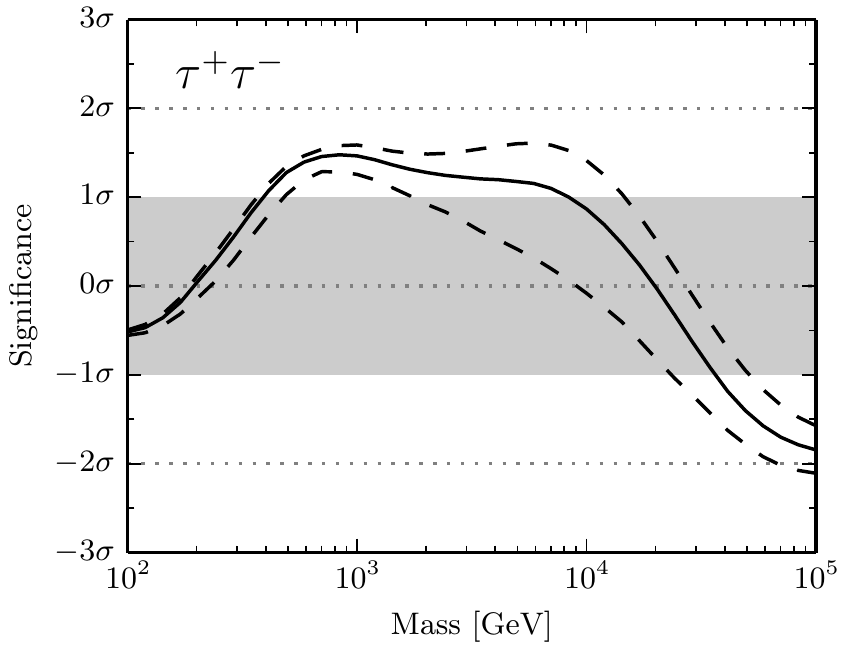}
	\put(20,15){\small\color{red} VERITAS - ICRC 2015}
	\end{overpic}
	\begin{overpic}[scale=0.55]{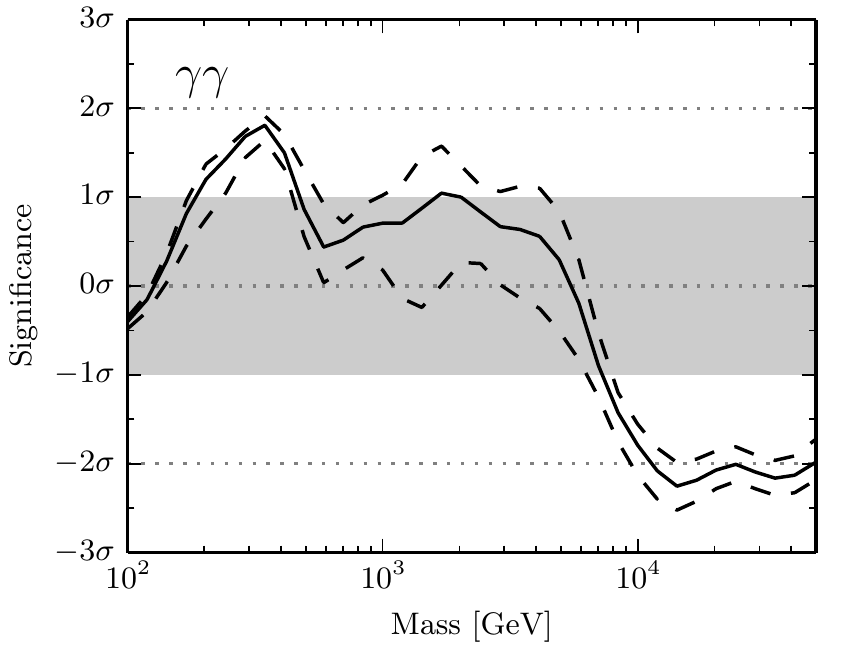}
	\put(20,15){\small\color{red} VERITAS - ICRC 2015}
	\end{overpic}
	\caption{Results of the search for dark matter annihilation of four of the dwarfs listed in Table 1 as a function of DM particle mass given annihilation into quarks ($b\bar{b}$, left), leptons ($\tau^{+}\tau^{-}$, center) or directly into $\gamma$-rays ($\gamma\gamma$, right). Detection significance of the DM hypothesis defined as the quantile of the probability distribution of the background-only model. Probability is converted to a sigma value by using the inverse CDF of a standard Gaussian. The dashed lines show the uncertainty in detection based on the DM density profile. The solid line is the median over the density profiles. } 
	\label{fig:search_indiv}
\end{figure}

\section{Limits on DM Annihilation Cross-section}

With the absence of a DM signal, we compute the limits on the annihilation cross section. The 95\% confidence limits are generated by performing the hypothesis test at every value of the cross section. The $\langle\sigma v\rangle$-space is divided in two regions where the hypothesis can and cannot be rejected at 95\% confidence. The hypothesis test is performed by asking, for a given value of $\langle\sigma v\rangle$, whether the probability that $T < T_{obs}$ is less than 5\%. The boundary between the two regions represents the 95\% upper limit to the cross section. 

Figure 3 shows the 95\% confidence limit obtained with the 216 hours of dSph data. Each panel uses a 100\% branching fraction into various Standard Model final states. The shaded band for the limits represents the 1$\sigma$ systematic uncertainty that exists because of imperfect knowledge of the dSph density profiles. Figure 4 shows the median observed DM limit for the $\tau^{+}\tau^{-}$ channel with 1$\sigma$ and 2$\sigma$ statistical uncertainties (the limit has a 68\% and 95\% chance to be within the red and green bands, respectively). Note that the observed limit does not exceed $\pm2\sigma$ of the expectation limit for any value of the DM mass. 

\begin{figure}
	\centering
	\includegraphics[scale=0.37]{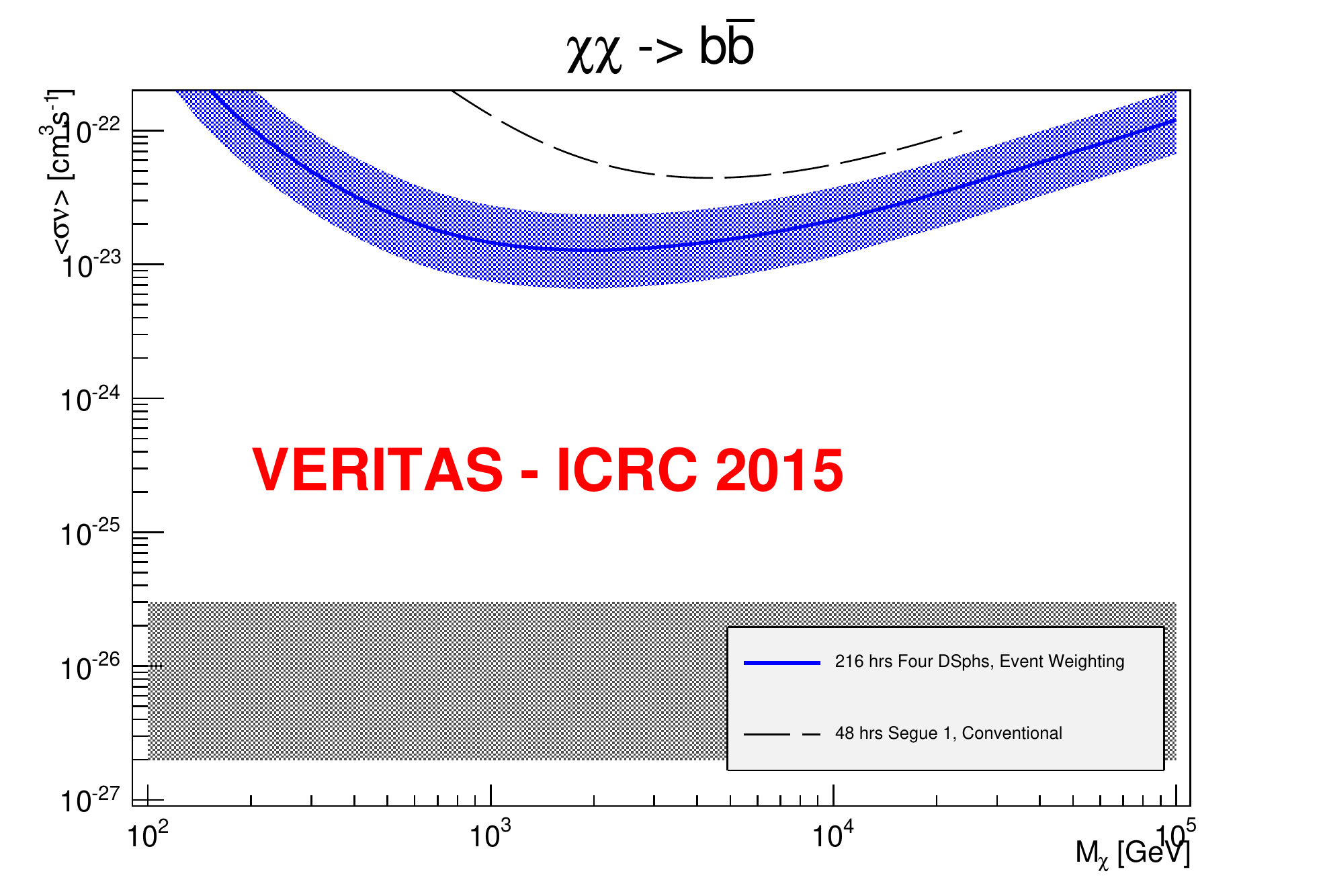}
	\includegraphics[scale=0.37]{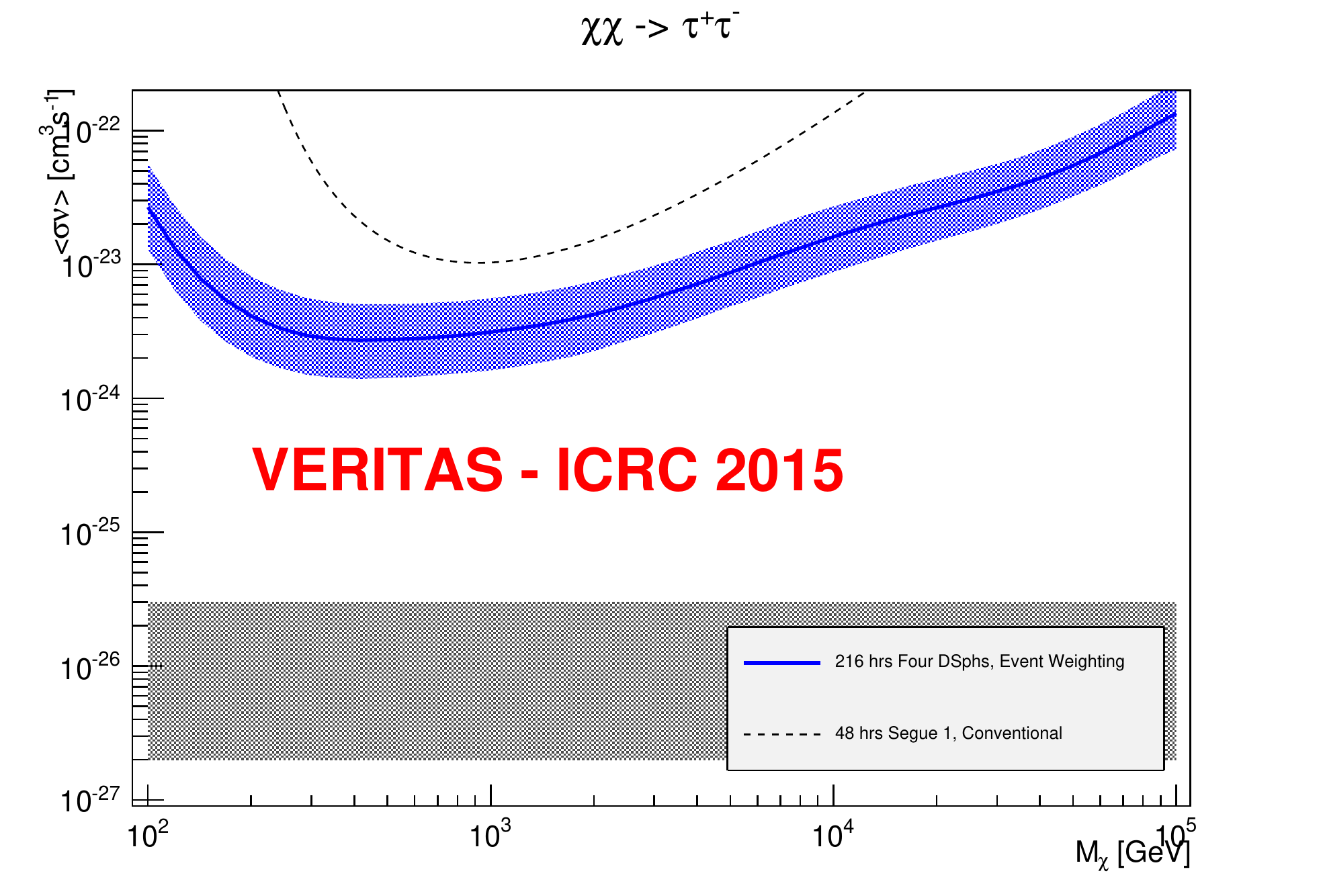}
	\includegraphics[scale=0.37]{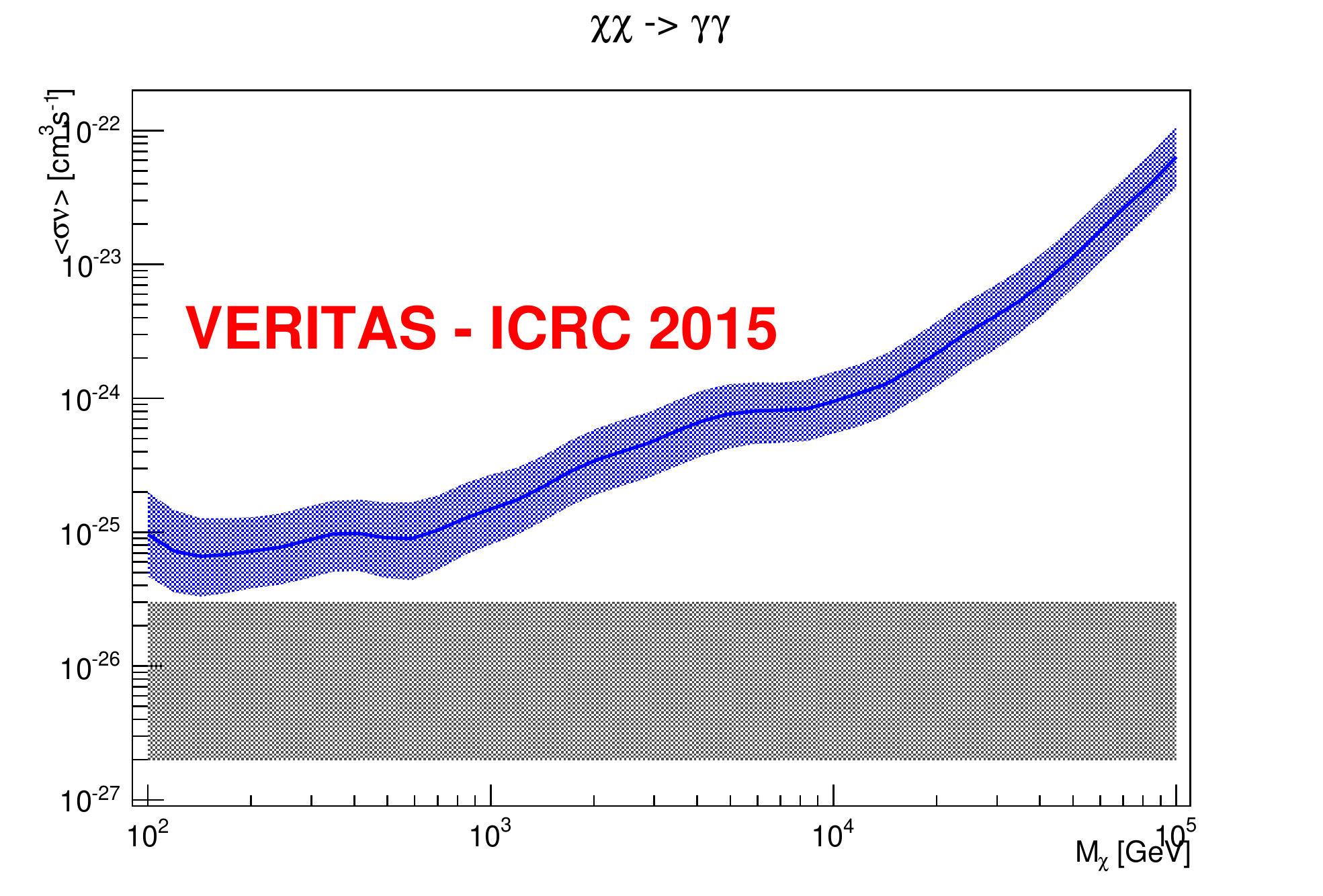}
	
	\caption{Preliminary observed cross section exclusion limits as a function of DM particle mass with 95\% confidence for annihilation channels $b\bar{b}$ (top left), $\tau^{+}\tau^{-}$ (top right) and directly to $\gamma\gamma$ (bottom). The blue bands indicate the 1$\sigma$ systematic uncertainty in the dark matter profile. In the cases of $b\bar{b}$ and $\tau^{+}\tau^{-}$, the cross section limits for the previously published 48 hour exposure VERITAS observations of Segue 1 are also shown \cite{2012PhRvD..85f2001A}. The Gray band represents a range of generic values for the annihilation cross-section in the case of thermally produced dark matter.}
	\label{fig:search_indiv}
\end{figure}

\begin{figure}
	\centering
	\includegraphics[scale=0.5]{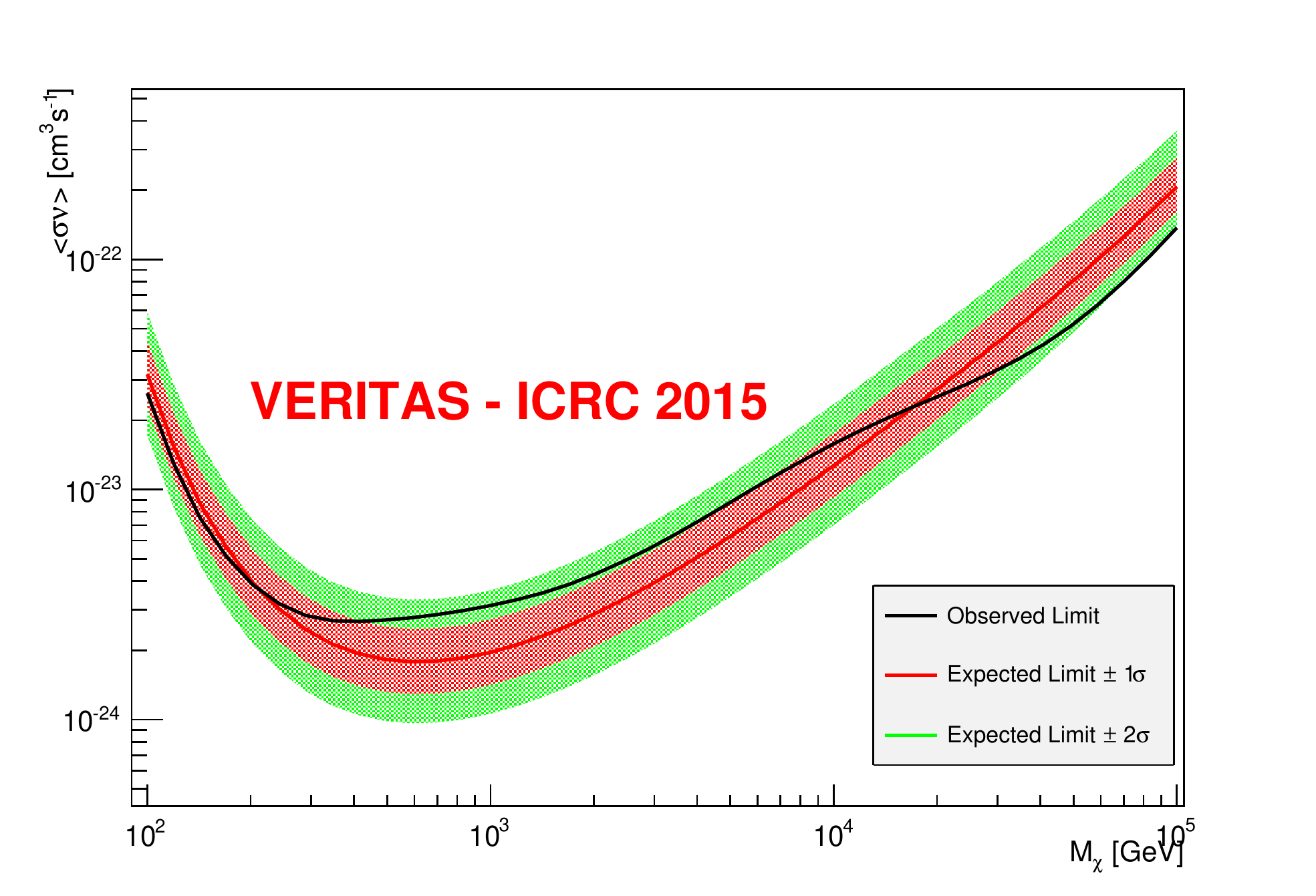}
	\caption{Preliminary observed cross section exclusion limits as a function of DM particle mass with 95\% confidence for DM annihilation to $\tau^{+}\tau^{-}$. The red solid line indicated the `expected' limit, generated from the background events. The red band and green bands represents the 1$\sigma$ and 2$\sigma$ probability of the limit being within those regions, respectively.  }
\end{figure}

\section{Conclusions}

Presented here for the first time are the results of using the event weighting method for IACT analysis, using information from several objects to perform a combined search and compute cross section limits.  No evidence of DM is found in 216 hours of combined dSph data, and limits of the annihilation cross section have been computed.  The event weighting method has the added benefit of using the VERITAS PSF and individual event reconstruction position in addition to using the individual energies of each event, leading to more accurate cross section limits. The combination of softer cuts used for the analysis, roughly twice the exposure for Segue and the event weighting method leads to much more constraining cross section limits than our previous Segue 1 analysis \cite{2012PhRvD..85f2001A} as shown in Figure 4. This is also the first time that a line search (DM annihilation directly to $\gamma$ rays) has been performed with VERITAS data.

\section{Acknowledgements}
This research is supported by grants from the U.S. Department of Energy Office of Science, 
the U.S. National Science Foundation and the Smithsonian Institution, and by NSERC in Canada. We acknowledge the excellent
work of the technical support staff at the Fred Lawrence Whipple Observatory and at the collaborating 
institutions in the construction and operation of the instrument. The VERITAS Collaboration is grateful to Trevor Weekes for his seminal contributions and leadership in the field of VHE gamma-ray astrophysics, which made this study possible.

\end{document}